\documentclass[12pt]{article}
\usepackage{amsmath}
\usepackage[dvips]{graphics}
\usepackage{latexsym}

\DeclareFontFamily{OT1}{rsfs}{}
\DeclareFontShape{OT1}{rsfs}{m}{n}{ <-7> rsfs5 <7-10> rsfs7 <10->
rsfs10}{} \DeclareMathAlphabet{\mycal}{OT1}{rsfs}{m}{n}
\def\scri{{\mycal I}}

\begin{document}
\newcommand{\bea}{\begin{eqnarray*}}
\newcommand{\eea}{\end{eqnarray*}}
\newcommand{\bean}{\begin{eqnarray}}
\newcommand{\eean}{\end{eqnarray}}
\newcommand{\eq}[1]{(\ref{#1})} 

\newcommand{\tri}{\delta}
\newcommand{\grad}{\nabla}
\newcommand{\pa}{\partial}
\newcommand{\bm}[1]{\mbox{\boldmath $#1$}}

\newcommand{\om}{\omega}
\newcommand{\omo}{\omega_0}
\newcommand{\ep}{\epsilon}
\newcommand{\nonu}{\nonumber}
\newcommand{\scrip}{\scri^{+}}
\newcommand{\hp}{{\cal H^+}}

\title{Late-time particle creation from gravitational collapse to an 
extremal Reissner-Nordstr\"om black hole}     
\author{Sijie Gao \\
Department of Physics \\   
University of Chicago \\
5640 S. Ellis Avenue \\
Chicago, Illinois 60637-1433, USA\\
and  Centro Multidisciplinar de Astrof\'{\i}sica - CENTRA,\\
Departamento de F\'{\i}sica, Instituto Superior T\'ecnico,\\
Av. Rovisco Pais 1, 1049-001 Lisboa, Portugal}

\maketitle

\begin{abstract}
We investigate the late-time behavior of particle creation from an
extremal Reissner-Nordstrom (RN) black hole formed by gravitational
collapse. We calculate explicitly the particle flux associated with
a massless scalar field at late times after the collapse. Our result
shows that the expected number of particles in any wave packet
spontaneously created from the ``in'' vacuum state approaches  zero
faster than any inverse power of time. This result confirms the traditional
belief that extremal black holes do not emit particles. We also
calculate the expectation
value of the stress energy tensor in a 1+1 RN black hole and show that it also
drops to zero at late times. Some comments on previous work by other
authors are  provided. 

\end{abstract}

\section{Introduction} 
As a quantum effect, particle production by black holes (Hawking
radiation) was
widely studied since the 1970s \cite{hawking}\cite{wald}\cite{wald2}. It
is well 
known that black holes emit particles with the same spectrum as a
blackbody 
with a temperature $T=\hbar \kappa/2 \pi k$, where $\kappa$ is
the surface gravity of the black hole and $k$ is the Boltzmann's
constant.  The Hawking radiation can be derived for a spacetime appropriate to
a collapsing body. At a late stage, the collapse settles down to a
stationary black hole.  Since spacetime is asymptotically flat at  both
past infinity  $\scri^-$ and  future infinity $\scri^+$, we can define the
``in''  and ``out'' vacuum states, respectively. If the two vacuum
states are distinct from one another, particles will be detected at future
infinity when the initial state is in ``in'' vacuum. If such creation
takes place at a steady rate at late times, it indicates that those
particles are 
produced from the stationary black hole instead of from the collapse phase. 

 Although the standard derivations of Hawking radiation only
deal with nonextremal black holes,  it is generally accepted that
extremal black holes have zero temperature and  consequently no particles
are created. However,  Liberati, {\em et al.} \cite{three} pointed out that
the generalization to
extremal black 
holes from nonextremal black holes is not trivial. One important
difference is that  the Kruskal transformation for non-extremal black
holes, which plays a crucial role in  computing the particle creation,
breaks down for extremal black holes. 

The arguments in \cite{three} are
reviewed briefly here as follows. Start with the usual form of the
Reissner-Nordstr\"om (RN)
geometry with parameters $Q$ and $M$
\bean
ds^2=-\left(1-\frac{2M}{r}+\frac{Q^2}{r^2}\right)dt^2
+\left(1-\frac{2M}{r}+\frac{Q^2}{r^2}\right)^{-1}dr^2+r^2d\Omega ^2,
\label{rn}
\eean 
where $d\Omega^2$ is the metric on the unit sphere. The tortoise
coordinate $r_*(Q,M)$ is given by
\bean
r_*(Q,M)=\int \frac{dr}{1-2M/r+Q^2/r^2}. \label{totg}
\eean
In the nonextremal case  $|Q|< M$, 
\bean
r_*(Q,M)=r+\frac{1}{2\sqrt{M^2-Q^2}}\left[r^2_+\ln(r-r_+)-r^2_-\ln(r-r_-)
\right]  \label{toqn}
\eean
where $r_{\pm}=M\pm\sqrt{M^2-Q^2}$. Define the retarded time $u$  and
advanced time $v$ as 
\bean
u&=&t-r_*,  \nonumber  \\
v &=& t+r_*. 
\eean
The well-known Kruskal transformation for the nonextremal case is
\bean
U&=&-e^{-\kappa u} \leftrightarrow u=-\frac{1}{\kappa}\ln(-U),\label{ku}  \\
V&=&e^{\kappa v} \leftrightarrow v=\frac{1}{\kappa}\ln(V), \label{kv}
\eean
where $U$ and $V$  are regular across the past and future horizons of
the extended spacetime. In the extremal
case $|Q|=M$, the 
right-hand side of Eq. \eq{toqn} appears to yield the indeterminate form
$0/0$. This can be fixed by setting $|Q|=M$ in Eq. \eq{totg} before
integrating. Thus,
\bean
r_*(M,M)=r+2 M \left(\ln (r-M)-\frac{M}{2(r-M)}\right). \label{rse}
\eean
Since $\kappa=0$ when $|Q|=M$, the Kruskal transformation \eq{ku} and
\eq{kv} breaks down for the extremal case. A generalization of the
Kruskal transformation to the extremal RN black hole is \cite{three}
\bean
u&=&-4M\left(\ln(-U)+\frac{M}{2 U}\right),  \label{gku}  \\
v&=&4M\left(\ln(V)-\frac{M}{2 V}\right).  \label{gkv}
\eean
We shall show, in section 2.2, that Eq. \eq{gku} defines a smooth
extension. However,  Liberati, {\em et al.} \cite{three} actually used a
simplified extension
\bean
u&=&-\frac{2M^2}{ U},  \label{sgku} 
\eean
which, as we will show later, is not a smooth extension (Eq. \eq{sgku} is
essentially the same extension introduced by Lake \cite{lake}). 
By using this extension,  Liberati, {\em et al.} \cite{three} calculated the
Bogoliubov 
coefficients associated with plane wave solutions of a massless scalar
field in a two-dimensional Minkowski spacetime with a moving mirror
(serving as a timelike boundary) which is physically equivalent
to  a (1+1)-dimensional model of an extremal RN spacetime formed from a
collapsing star. The
result shows that the 
Bogoliubov coefficients are nonzero, indicating that particles are
created in the late stages of collapse. Further calculations in
\cite{three} also show that the expectation value of the
stress-energy-momentum tensor is zero and its variance vanishes as a
power law at late times. The authors thereby claim that the extremal
black hole does not behave as a thermal object and cannot be regarded as
the thermodynamic limit of a nonextremal black hole. 

However, the major deficiency in the analysis of \cite{three} is the
use of unnormalized plane-wave solutions. These kinds of solutions have
been used for nonextremal cases \cite{davis}\cite{m1}\cite{dfp}. The
Bogoliubov coefficients 
$\beta_{\om\om'}$ in \cite{three} have the form
\bean
\beta_{\om\om'}\sim \sqrt{\frac{\om'}{\om}}\int_0^\infty  e^{-i\om'
  v+\frac{i a \om}{v}}dv. \label{bif} 
\eean
The integrand is oscillated with constant amplitude. So the integral is
not well-defined. The result 
\bean
\beta_{\om\om'}\sim K_1(2\sqrt{a\om\om'}) \label{bbs}
\eean
given in \cite{three}, which  was originally calculated by Davis and Fulling
\cite{dfp}, was obtained by Wick rotation, i.e., integrating along  the
imaginary axis. But this Wick rotation is unjustified 
since the integrand does not fall off at large radius on the complex
plane. Since the spectrum of particle number created from the vacuum
is 
\bean
<N_{\om} >=\int_0^{\infty}|\beta_{\om\om'}|^2 d\om' \label{pnb}
\eean
and $K_1(z)\sim 1/z$ for $z\rightarrow 0$ \cite{bes}, the number of
particle is divergent.  The authors interpret the infinity as  an
accumulation after an infinite time. The Kruskal extension \eq{sgku}
was used in deriving Eq. \eq{bif}. If we use the smooth extension \eq{gku}
instead, the Bogoliubov coefficients would be 
\bean
\beta_{\om\om'}\sim \sqrt{\frac{\om'}{\om}}\int_0^\infty  e^{-i\om'
  v+\frac{i a \om}{v}}e^{-4iM\om\ln(v)}  dv \label{bigf} 
\eean
and by using the same Wick rotation (also unjustified), it follows that
\bean
\beta_{\om\om'}\sim
\om'^{2iM\om}K_{1-4iM\om}(2\sqrt{a\om\om'}).\label{sme}
\eean
For small $z$, $K_{1-4iM\om}(z)
\approx\frac{1}{2}\Gamma(1-4iM\om)(\frac{z}{2})^{-1+4iM\om}$.
Therefore, the number of particle in Eq.\eq{pnb} is still infinite. 

To clarify this issue, our main calculation focuses on wave-packet
solutions with unit Klein-Gordon norm. The wave packets $P_{n\ep\omo lm}$
we will construct are made up of frequencies within $\ep$ of
$\omo$. They are peaked around the retarded time $u=2\pi n$ and have a
time spread $\sim 2\pi/\ep$. The created particle number, $N_{n\ep}(\omo)$,
associated with the wave packet has a direct physical interpretation:
$N_{n\ep}(\omo) $ is proportional to the counts of a particle detector
sensitive only 
to frequencies within $\ep$ of $\omo$ and angular dependence $Y_{lm}$
which is turned on for a time interval $2\pi/\ep$ at time $u=2\pi
n$. Our calculation shows that for fixed $\omo$, $\ep$, $l$ and $m$,
$N_{n\ep}(\omo)$ drops off to zero faster than any inverse power of $n$.
Therefore, 
the traditional belief that extremal black holes do not 
emit particles is confirmed. Furthermore, if we sum $N_{n\ep}(\omo)$
over the integers 
$n$, we still get a finite
result. This indicates that, even after an infinite time, the
accumulation of particles for a certain frequency is still finite. This
contradicts the infinite result in \cite{three}. Our calculation is
independent of choice of a specific type of wave packet provided that
its Fourier transform is a $C^{\infty}$ function with compact support
on purely  positive frequencies. As explained in section 2.4, we also
conjecture that our result is independent of the details of the collapse.

Note that two independent errors were made in \cite{three}. First, the
nonsmooth Kruskal extension \eq{sgku} was used rather than the smooth
Kruskal extension \eq{gku}. If Eq.\eq{sgku} were used in the wave-packet
method, the Wick rotation used in calculating the negative frequency
part of the wave packet at the past infinity would not be
justified(See footnote 1). Second, unnormalizable plane waves were used 
rather than normalized wave packets. Even if the Kruskal extension \eq{gku}
had been used, the use of un-normalizable plane waves would have
resulted in the prediction of an infinite number of particles.

We also calculate the expectation value of stress-energy tensor 
$<T_{uu}>$ related to the extension \eq{nkru} and find that it
drops to zero as $\frac{1}{u^3}$. This conclusion is proved to be
independent of the details of the collapse.  From the particle flux in a wave
packet,  we find that $\int_0^{\infty} N_{n\ep}(\omo)(\omo)\omo d\omo$
drops as fast as or faster than  $1/u$, which is not in contradiction with the
$\frac{1}{u^3}$ decay rate.

Our calculation follows the similar steps to \cite{hawking}. We focus on
a massless scalar field on an extremal RN black-hole spacetime  which is formed
from a collapsing star.   We start 
by constructing a positive frequency (relative to retarded time $u$)
wave packet $P_{n\ep\omo lm}$ at 
future infinity  $\scri^+$. By  using the geometrical optics approximation, we
propagate  the solution back to the past infinity $\scri^-$. The 
particle number in this mode can be obtained by computing the
Klein-Gordon norm of the negative frequency (relative to advanced time
$v$) part of  $P_{n\ep\omo lm}$ at
$\scri^-$.

\section{Calculation of particle creation}
\subsection{Construction of the wave packets at future infinity}
Our purpose in this subsection is to construct positive frequency wave packets 
at future infinity $\scri^+$.
We start with the massless Klein-Gordon equation   $\Box\phi=0$.  In the
region outside the 
collapsing matter, the 
spacetime is described by the extremal RN metric \eq{rn}. 
Write
$\phi=r^{-1}F(r,t)Y_{lm}(\theta,\phi)$, where $Y_{lm}(\theta,\phi)$ is a
spherical harmonic. Then outside the collapsing
matter, 
$\Box\phi=0$  yields 
\bean
\frac{\pa^2 F}{\pa t^2}-\frac{\pa^2 F}{\pa
r_*^2}+V(r)F=0, \label{ef}
\eean
where
\bean
V(r)=-\frac{(M-r)^2}{r^6}(2 M^2-2Mr+l(l+1)r^2). \label{pv}
\eean
Furthermore, assume $F(r,t)=g(r_*)e^{i\omo u}$, where  $u=t-r_*$. Then
Eq. \eq{ef} becomes 
\bean
2i\omo g'(r_*)-g''(r_*)+g(r_*)V(r)=0. \label{neg}
\eean

As $r\rightarrow \infty$, $r_*/r\rightarrow 1$ and $V(r)\rightarrow 0$. So
$g(r_*)$ approaches a constant and 
\bean
P_{\om_0lm}=\frac{1}{r}e^{i\omo u}Y_{lm}(\theta,\phi) \label{sps}
\eean
 is a solution  near $\scri^+$. A wave packet with frequencies around
$\omo$ and centered on retarded time $u=2n\pi$   can be
constructed by superposing the above spherical waves as
\bean
P_{n\ep\om_0
lm}=\frac{1}{r}z_{\om_0 n}(u)Y_{lm}(\theta,\phi), \label{ssl}
\eean
where
\bean
z_{\om_0 n}(u)=A\frac{1}{\sqrt{2\pi}} \int_{-\infty}^{\infty}
f\left(\frac{\om-\omo}{\ep}\right)e^{-i 
2 n \pi \om}e^{i\om u} d\om \label{ft}
\eean
where  $A$ is a normalization constant and  $ f(x)$ is a real
$C^\infty$ function 
with compact support in $x\in [-1,1]$.
To guarantee $P_{n\ep\omo lm}$ has positive frequencies near $\omo>0$, we
require $0<\ep \ll\omo$. Let
$\frac{\om-\omo}{\ep}=\tilde\om$ and $u-2 
n\pi=\tilde u$. Then we may rewrite Eq. \eq{ft} as
\bean
z_{\om_0 n}(u)=A\ep e^{i\omo \tilde u} \hat f(\ep\tilde u),\label{ftt} 
\eean
where
\bean
\hat f(x)= \frac{1}{\sqrt{2\pi}}\int_{-\infty}^{\infty}  f(\tilde
\om)e^{i \tilde\om x} d\tilde\om.  \label{fhat}
\eean

The normalization constant $A$ is determined by the Klein-Gordon inner
product \cite{gr}
\bean
(P,P)_{KG}=i\int_\Sigma (\bar P\grad_a P-P\grad_a \bar P )n^a dV=1 \label{kg}
\eean
Taking $\Sigma$ to be  $\scri^+$, Eq. \eq{kg} becomes
\bean
-i\int \left(\bar P\frac{\pa P}{\pa u}-P \frac{\pa \bar P}{\pa u} \right)r^2
d\Omega du=1 \label{ppi}
\eean
Substituting \eq{ssl} into \eq{ppi} gives
\bean
-i \int_{-\infty}^{\infty} \left[  \bar z(u)z'(u)-z \bar z'(u)\right]
du=1, \label{zz} 
\eean
where we have omitted the subscripts of $z_{\omo n}$. Straightforward
calculation from Eq. \eq{ftt} gives
\bean
&&\bar z(u)z'(u)-z \bar z'(u) \nonu \\
&=& i 2\omo A^2\ep^2 \left| \hat f(\ep\tilde u)\right|^2 \nonumber \\
&+&\frac{1}{2\pi}
i2A^2\ep^3 Re\left[  \left(\int_{-\infty}^{\infty} f(\tilde 
\om)e^{i\ep\tilde u \tilde\om } d\tilde\om \right)\left(\int_{-\infty}^{\infty}
 f(\tilde \om)\tilde\om e^{-i\ep\tilde u \tilde\om } d\tilde\om \right)
\right] \label{zuz} 
\eean
The solution for $A$ is found to be
\bean
A=\frac{1}{\sqrt{\beta \ep\omo+\gamma\ep^2}},  
\eean 
where $\beta$ and $\gamma$ are integral constants defined by
\bean
\beta&=&2\int_{-\infty}^{\infty}dx  
\left|\hat f(x) \right|^2 \nonu \\
&=&2\int_{-\infty}^{\infty} |f(\tilde\om)|^2 d \tilde\om \\
\gamma &=&2\frac{1}{2\pi} Re\left[ \int_{-\infty}^{\infty}dx
\left(\int_{-\infty}^{\infty}e^{ix\tilde\om} f(\tilde\om)d
\tilde\om\right)\left(\int_{-\infty}^{\infty}e^{-ix\tilde\om}
f(\tilde\om)\tilde\om 
d\tilde\om\right)  \right] \nonu \\
&=&2\int_{-\infty}^{\infty}\tilde\om | f(\tilde\om)|^2 d \tilde\om
\eean
Therefore, Eq. \eq{ftt} becomes
\bean
z(u)=\frac{\sqrt\ep}{\sqrt{\beta \omo+\gamma\ep}}  e^{i\omo \tilde u}
\hat f(\ep\tilde u).       \label{zu}
\eean

\subsection{Kruskal coordinates}
Kruskal coordinates will play an important role in our following
calculation. Specifically,  we seek a coordinate $U$ which is a
smooth function of $u$ outside of the black horizon and covers a
neighborhood of the horizon with $U=0$ on the horizon such that the
metric in the coordinates $(U,v,\theta,\phi)$ is smooth on the horizon.
Note that $r$
is a smooth function (it is easy to check that $r$ is an affine
parameter of an incoming null geodesic). Thus, along an ingoing null
geodesic with constant $v$, an affine parameter $U$ can be taken as
$U=-(r-M)$ and $U(u)$ is obtained from Eq. \eq{rse} by using
$r_*=\frac{1}{2}(v-u)$. Therefore, we have
\bean
u=2U-4M\left(\ln(-U)+\frac{M}{2 U}\right)+d,  \label{nkru}  
\eean
where $d$ is a constant. Without loss of generality, we may assume
$d=0$. The coordinate $U$ defined along
the ingoing null geodesic can be ``carried'' away by outgoing null
geodesics. For a smooth two-dimensional spacetime defined by $(U,v)$,
there exist smooth coordinates $(\hat U,\hat V)$ such that the metric
takes the form
\bean
ds^2=-\Omega^2(\hat U,\hat V)d\hat U d\hat V, 
\eean
where $\hat U=\hat U(U)$. Since $\hat U$ must be  a smooth function of $U$
with nonzero first derivative along the null ingoing geodesic where
$U$ is an affine parameter,  $\hat U=\hat U(U)$ defines a smooth
coordinate transformation.  Therefore, we have constructed a smooth
Kruskal extension $U$. Next, we
wish to show that Eq. \eq{gku} also defines a smooth extension. To
distinguish, we rewrite $U$ in Eq. \eq{gku} as $U'$. So we have, from
Eqs. \eq{gku} and \eq{nkru} 
\bean
-4M\left(\ln(-U')+\frac{M}{2 U'}\right)=2U-4M\left(\ln(-U)+\frac{M}{2
  U}\right) \label{upu}
\eean
Since $U'$ is obviously smooth outside the black hole, we only need to
show that $U'$ is a smooth coordinate in a neighborhood of the
horizon, i.e., $U'$ 
is a smooth function of $U$ and $\frac{dU'}{dU}\neq 0$ around
$U=0$. Let $U'=[1+U h(U)]$ and 
substitute into  Eq. \eq{upu}. Differentiating both sides of \eq{upu}
with respect to $U$, we obtain
\bean
\frac{dh}{dU}=\frac{(1+(M+U)h)^2}{M(M-2U-2U^2 h)}\label{hpu}
\eean
It is easy to check that the right-hand side of Eq. \eq{hpu} is a smooth
function of $(U,h)$ at $U=h=0$. Therefore, according to the theory of
ordinary differential equations, there exists a unique smooth solution
to $h$ around $U=0$ with $h(0)=0$. It is also easy to check that
$\frac{dU'}{dU}\neq 0$  around $U=0$. Therefore, $U'$ is a smooth
extension, i.e., the $U$ defined by \eq{gku} is a smooth coordinate. 
Now we show  that $U$ defined by \eq{sgku} is not a smooth coordinate
on the horizon. Rewrite  $U$ in \eq{sgku} as $U'$. Then, from \eq{gku}
and \eq{sgku}, we express $U'$ as 
\bean
U'=\frac{M U}{M+2U\ln(-U)} \label{uns}
\eean
It is straightforward to show that $\frac{d^2U'}{dU^2}$ is divergent
at $U=0$. Therefore, the extension defined by Eq. \eq{sgku} is not smooth.

\begin{figure}[t]
\resizebox{\textwidth}{!}
{\includegraphics[-2.5in,0in][9.5in,6in]{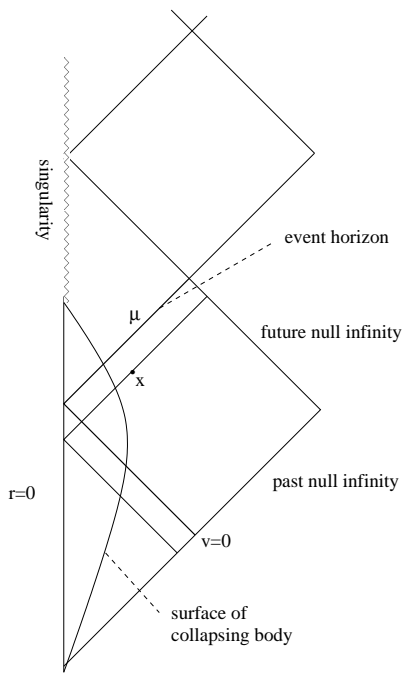}}
\caption{The Penrose diagram of a spherically symmetric collapsing body
producing an extremal RN black hole \cite{boulware}. (The body is shown
collapsing to a singularity at $r=0$, but it might instead re-expand
into the new asymptotically flat region; see \cite{boulware} for further
discussion of the behavior of charged shells. The behavior of the body
after it crosses the event horizon is, of course, not relevant for our
analysis.) }
\end{figure}

\subsection{Geometrical optics approximation}
To calculate the particle creation rates at late times, we need to
propagate the wave packet \eq{ssl} backward
from $\scri^+$ to $\scri^-$. For simplicity, we first investigate
the propagation of solution \eq{sps}; later, the propagation of
Eq. \eq{ssl} can be easily obtained by superposition.  A part of the wave
\eq{sps}  will be scattered by the static Schwarzschild field outside
the collapsing body 
and will end up on $\scri^-$ with the same frequency \cite{hawking} and
will not contribute to particle creation. We are interested in the
remaining part which will propagate through the center
of the collapsing star, eventually emerging to $\scri^-$. Consider the
solution \eq{sps} propagating to a point $x$, which is  very near the
future event horizon ${\cal H^+}$ and outside the collapsing body (see
Fig. 1). The solution near $x$
has a form similar to its form on $\scri^+$,
\bean
P_{\om_0lm}(x)\sim t(\omo)e^{i\omo u}Y_{lm}(\theta,\phi), \label{sx}
\eean
where $t(\omo)$ is the transmission amplitude describing the
fraction of the wave that enters the collapsing body. 
Note that near the horizon, the effective frequency will be arbitrarily
large \cite{gr}. So the amplitude of Eq. \eq{sx} changes much slower than
the phase. Consequently,  the geometrical optics approximation becomes
valid for the propagation from $x$ back to $\scri^-$. So the wave takes
the form 
\bean
P_{\omo lm}=g(t,r)e^{i\omo S}Y_{lm}(\theta,\phi), \label{wap}
\eean
where $S$ is called the phase of the wave. Each surface of constant $S$
is a null 
hypersurface \cite{gr} and consequently $k^a\equiv
\grad^a S$  is the tangent to the null geodesics propagating in
the radial direction. If we follow a light ray backward in time, it will
pass through the center of the star and propagate to $\scri^-$. We fix
$(\theta,\phi)$ and  consider the family of radial  null geodesics
$\gamma_\chi(\lambda)$ such that 
for each $\chi$,  $\gamma_\chi(\lambda)$ represents a null geodesic with
parameter $\lambda$ sent from $\scri^+$ radially to the collapsing
star. So all geodesics in $\gamma _\chi(\lambda)$ have the same 
path in space but they pass through  the center of the star at
different times. The limiting null geodesic  in this family lies  on
the future horizon.  Set $\chi=0$ for this geodesic and denote it by
$\mu$, i.e., $\mu=\gamma_0(\lambda)$. Let $x$ be an event lying just
outside the 
horizon (see Fig. 1). According to 
geometrical optics \cite{gr}, $S=u(x)$ along a null geodesic.
To find out the explicit form of $S$, let $v$
be the Killing/affine  parameter coordinate at past null infinity and
$v=0$ correspond to the light ray on the horizon. For a fixed
collapse, define $v$ on the spacetime by propagation from past null
infinity along radial null geodesics. Let $U$ be a smooth Kruskal
extension such that  it is a constant along each outgoing null
geodesic and $U=0$ on the horizon. Therefore, a function $v(U)$ can be
constructed from those radial null geodesics with $v(0)=0$. The exact
form of $v(U)$ should be solved from the geodesic equation which
depends on the details of the collapse. However, no matter what the
details of the collapse are, the corresponding spacetime must be
smooth. Consequently, the geodesic equation is a smooth equation and
thereby $v(U)$ is a smooth function for $U\leq 0$. Equivalently, each
smooth Kruskal extension $U$ should correspond to some smooth collapse
spacetime for which the propagation of radial null geodesics from
future infinity to past null infinity is given by $v = U(u)$. Thus,
for the Kruskal extension defined by Eq. \eq{gku}, we have  
\bean
u=S(v)=-4 M \ln (-v)-\frac{2 M^2}{v}. \label{sv}
\eean

There is a close analog between four-dimensional spherical collapse and
two-dimensional Minkowski spacetime with a moving mirror. The physical
relations have been widely discussed in previous literature,
e.g., \cite{davis}, \cite{m1} and \cite{dfp}. We shall only illustrate
the mathematical correspondence between a spherical collapse and a
mirror trajectory. In a two-dimensional Minkowski spacetime with
double-null coordinates $(u,v)$, a moving mirror servers as boundary of
the spacetime. If a left-moving light ray with constant $v$ is
reflected by the mirror, it then becomes right-moving with constant
$u$. The relation between $u$ and $v$ is uniquely determined by the
coordinates, $(u,v)$, of the reflecting point on the mirror. Thus, the
mirror trajectory $u=u(v)$ shows how a light ray  propagates after
reflection. Let the left-moving light ray correspond to an ingoing light
ray in a spherical collapse and the right-moving light ray correspond to an
outgoing one. Then we see that the mirror plays the role of the
origin of spherical coordinates. The trajectory associated with the
collapse above is simply Eq. \eq{sv}. Such a relation will be used later to
calculate the energy flux.

The amplitude in Eq. \eq{wap} near $\scri^-$  can be calculated by substituting
Eq. \eq{wap} into $\Box\phi=0$. After neglecting the $\Box g$ term, which
is supposed to be small in the  geometrical optics approximation, we obtain
\bean
2 k^a\grad_a g+g\grad_a k^a=0. \label{dap}
\eean
Note that $\grad_a k^a$ is the expansion of the congruence of radial
null geodesics, which is equal to
\cite{wald2} 
\bean
\frac{1}{A}\frac{dA}{d\lambda}, \label{ared}
\eean
where $\lambda$ is the parameter of $k^a$ and $A$ is the cross-sectional
area element. Since all geodesics represented by $k^a$ point radially
and the spacetime is  spherically symmetric, $A\propto r^2$. Hence,
Eq. \eq{dap}
gives 
\bean
\frac{d\ln(gr)}{d\lambda}=0. \label{glc}
\eean 
Namely, $gr$ is a constant along each null geodesic. So $gr$ is
proportional to $t(\omo)$ in Eq. \eq{sx}. Finally we get the solution near the
past infinity,
\bean
P_{\omo lm}\sim 
\left\{
\begin{array}{ll}
t(\omo)\frac{1}{r}e^{i\omo S(v)}Y_{lm}(\theta,\phi), & v<0 \\
0, & v>0, 
\end{array}
\right.
\label{ssp} 
\eean
where $S(v)$ is given in Eq. \eq{sv}. Next, superpose solutions the same
way as we did on $\scri^+$ (refer to \eq{ft}) and assume that $t(\om)$
varies negligibly over the frequency interval $2\ep$. Then, we obtain the
wave packet at $\scri^-$,
\bean
P_{n\ep\omo lm}\sim 
\left\{
\begin{array}{ll}
t(\omo)\frac{1}{r}z_p(v)Y_{lm}(\theta,\phi), & v<0 \\
0, & v>0, 
\end{array}
\right. \label{wpp}
\eean
where 
\bean
z_p(v)&\equiv& z_{\omo n}(S(v)) \nonumber \\
&=&\frac{\sqrt\ep}{\sqrt{\beta \omo+\gamma\ep}} e^{-i\omo 2
n\pi}e^{i\omo[-4 M \ln(-v)-\frac{2  M^2}{v}]}\hat
f(\ep\tilde u)  \label{zv}
\eean
and
\bean
\tilde u=-4 M \ln (-v)-\frac{2 M^2}{v}-2n\pi. \label{utv}
\eean

\subsection{Calculation of particle creation}
We shall show that the particle creation rate for each
mode with fixed $\omo$, $\ep$, $l$ and $m$ will drop off to zero at
sufficiently late times. So in this subsection, we treat $\omo$ and $\ep$
as fixed  and consider the limit where  $n$ is allowed to become arbitrarily
large. The 
expected number of particles spontaneously created  in the  state
represented by a wave packet is given by \cite{gr}
\bean
N_{n\ep}(\omo)=(P^-,P^-)_{KG}, \label{pcr}
\eean
where $P^-$ represents the negative frequency part of the solution
\eq{wpp}. The negative frequency is with respect to $v$. Since  the only
$v$ dependence in Eq. \eq{wpp} is $z_p(v)$, after 
integrating on $\scri^-$, \eq{pcr} reduces to 
\bean
N_{n\ep}(\omo)=|t(\omo)|^2\int_0^\infty |\hat z(\om')|^2\om' d\om', \label{npc}
\eean
where 
\bean
\hat z(\om')&=&\int_{-\infty}^{0}e^{i\om'v}z_p(v)dv \label{dzp} \\
&=&\frac{\sqrt\ep}{\sqrt{\beta
\omo+\gamma\ep}} \int_{-\infty}^{0}e^{i\om'v}  e^{-i\omo 2 
n\pi}e^{i\omo[-4 M \ln(-v/\alpha)-\frac{2 \alpha M^2}{v}]}\hat
f(\ep\tilde u) dv
\label{dzo}
\eean
is the amplitude of the negative frequency part of $z_p(v)$. Note that in
$z_p(v)$, $v$ is always multiplied by an undetermined factor
$1/\alpha$.  By a simple rescaling, we see immediately that
$N_{n\ep}(\omo)$ is independent of the choice of $\alpha$. So without
loss of generality, we set $\alpha=1$ from now on. The
difficulty in  evaluating Eq. \eq{dzo} is the oscillation in the integrand. We
wish to eliminate this oscillation by a Wick rotation. We shall show that
the integral in Eq. \eq{dzo} can be performed along the positive imaginary
axis in the complex $v$ plane. 
To justify the
rotation, we need to use the following theorem, which corresponds to one
direction 
of the Paley-Wiener theorem \cite{theo}.
\newtheorem{thm}{Theorem}
\begin{thm}
Let $f: {\mathrm{I\!R}} \rightarrow {\mathrm{I\!R}}$ be a $C^{\infty}$
function with support in 
$[-1,1]$. Then, the Fourier transform, $\hat f(\zeta)$, of $f$ is an
entire analytic function of $\zeta$ such that for all $k>0$, 
\bean
|\hat f(\zeta)|\leq\frac{C_k e^{|Im \zeta|}}{(1+|\zeta|)^k}   \label{ineq}
\eean
for all $\zeta\in {\bm C}$, where $C_k$ is a constant which
depends on $k$. 
\end{thm} 
We shall be interested in large $k$. Hence, $k$ is assumed to be large
in the rest of the paper. 
Applying this theorem to $\hat f(x)$ defined in \eq{fhat} with $x$ replaced
by $\ep \tilde u$, we find immediately
\bean
|\hat f(\ep \tilde u)|\leq\frac{C_k e^{|Im [\ep\tilde u]|}}{(1+|\ep\tilde
u|)^k} \label{fin}
\eean
Thus, in contrast to the integrals considered in \cite{three}, the
integral appearing in Eq. \eq{dzo} is convergent.
The theorem also tells us that $\hat f(\ep\tilde u)$ is an analytic
function of $\ep\tilde u$. Thus, the integrand of Eq. \eq{dzo} is  analytic
everywhere in the 
second quadrant except at the origin of the 
$v$-plane. However, we can choose a contour which goes around the origin
along a circle with infinitesimal radius in the first quadrant. In order
to apply Cauchy's integral theorem, we choose the closed contour as shown in
Fig. 2, where $C_1$ and $C_2$ are two circles with small and
large radii, respectively. We are going to show that the integration
in Eq. \eq{dzo} over the two 
circles is negligible. As for the small circle $C_1$,  we need to show
that  the
integrand is not divergent in the second quadrant near the origin. From
Eqs. \eq{utv} and \eq{fin}, it is easy to see that the only possible source
causing the divergence near the origin is $e^{|Im[\ep\frac{2\alpha
M^2}{v}]|}$.  However, this term is always suppressed by $e^{-i\frac{2\alpha
M^2\omo \om'}{v}}$ in Eq. \eq{zv} since $\omo>\ep$. Therefore, the
integral over 
 $C_1$  approaches zero. To deal with the integration over $C_2$,
we introduce the following lemma.
\newtheorem{lem}{Lemma}
\begin{lem}
If $F(z)$  satisfies $\lim_{|z|\rightarrow \infty} |F(z)|=0$ in the
second quadrant, then \\
$\int_{C_2}e^{icz} F(z)=0$ when the radius of $C_2$
approaches  infinity, where $c>0$ is a constant. 
\end{lem}
\begin{figure}[t]
\resizebox{\textwidth}{!}
{\includegraphics[-2.5in,0in][9.5in,6in]{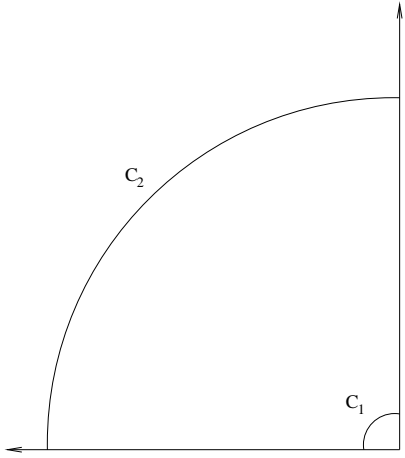}}
\caption{ Integration contour in the $v$-plane }
\end{figure}
The proof of the lemma is given in the appendix. From Eq. \eq{dzo} and
\eq{fin}, it is easy to see that the condition of the lemma is
satisfied.\footnote{Keeping the logarithm in the $\tilde u(v)$ is
essential to make the condition satisfied. Otherwise, $z_p(v)$  will not
drop to zero for large $|v|$. So the rotation does not apply to the
accelerated mirror case. }  Therefore, the integration over $C_2$
approaches zero. Thus, the integral in Eq. \eq{dzo} can be performed along
the  positive imaginary axis in the $v$-plane. Let
$x=-i\om'v$. Then the integral in Eq. \eq{dzo} corresponds to  the positive
real axis in the $x$-plane,
\bean
\hat z(\om')=\frac{\sqrt\ep e^{-i\omo 2 n\pi} }{-i\om'\sqrt{\beta
\omo+\gamma\ep}} 
\int_{0}^{ \infty} e^{-x} e^{-i\omo 4 M \ln(\frac{x}{i\om'})}e^{-\frac{2
M^2\omo \om'}{x}}\hat f(\ep\tilde u(x)) dx, \label{zox}        
\eean
where 
\bean
\tilde u(x)=-4 M\ln(\frac{x}{i\om'})+\frac{i 2 M^2\om'}{x}-2 n\pi. \label{ux}
\eean
Using the fact that $Im [\tilde u]=2 M
\pi+\frac{2 M^2\om'}{x}$ and $|e^{-i\omo 4M \ln(\frac{x}{i\om'})}|=
e^{-2M\pi\omo}$  when
$x$ is real, together with Eq. \eq{zox},  we have
\bean
|\hat
z(\om')|&\leq&\frac{\sqrt{\ep}C_k e^{-2M\pi
(\omo-\ep)}}{\om'\sqrt{\beta\omo+\gamma\ep}}\int_0^    
{\infty}\frac{e^{-x}e^{-\frac{2 M^2\om'(\omo-\ep)}{x}}}{(1+|\ep \tilde
u(x)|)^k} dx 
\label{oni}
\\
&\leq&\frac{\sqrt{\ep}C_k e^{-2M\pi (\omo-\ep)}
}{\om'\sqrt{\beta\omo+\gamma\ep}}\int_0^{\infty} 
\frac{e^{-x}e^{-\frac{2 M^2\om'(\omo-\ep)}{x}}}{|\ep \tilde u(x)|^k} dx.
\label{inex} 
\eean
To proceed, we first derive a lower bound for $|\tilde u(x)|$ at large $n$
. Start with  
\bean
|\tilde u(x)|^2=\left(2n\pi+4 M\ln(\frac{x}{\om'})\right)^2+\left(2 M
\pi+\frac{2 M^2\om'}{x} \right)^2.  \label{abu}
\eean
Let $y=\frac{x}{\om'}$ and define
\bean
h(y)\equiv |\tilde u(x)|^2=\left(2n\pi+4 M\ln y\right)^2+\left(2 M
\pi+\frac{2 M^2}{y} \right)^2
\eean
To find the minimum, we solve $h'(y)=0$, which gives
\bean
-\frac{1}{y^3}(8 M^4 +8 M^3 \pi y-16 M n \pi y^2 -32 M^2 y^2 \ln
y)=0. \label{sfm} 
\eean
Obviously, $y=0$ is not a solution where $h(y)$ achieves its minimum. 
When $n$ is a large number, the solution to \eq{sfm} must be at small
$y$. So  the approximate solution is
\bean
y_0=\frac{M^{3/2}}{\sqrt n\sqrt{2 \pi}}
\eean
and
\bean
h(y_0)=(2 n\pi)^2.
\eean
It is easy to check, by computing the second derivative, that $h(y_0)$ is a
minimum for large $n$.   So we have an important inequality
\bean
|\tilde u(x)|\geq 2 n \pi         \label{geqn}
\eean 
for large $n$. Then it follows immediately from Eq. \eq{inex} that
\bean
|\hat z(\om')|&\leq&\frac{\sqrt{\ep}C_k  e^{-2M\pi
(\omo-\ep)}}{\om'\sqrt{\beta\omo+\gamma\ep}}\frac{1}{\ep 
^k (2n\pi)^k}
 \int_0^{\infty}e^{-x}e^{-\frac{2 M^2\om'(\omo-\ep)}{x}}  dx. \label{zine}
\eean
However, this bound is not good enough for
all $\om'$ since we  must integrate $|\hat z(\om')|^2\om'$ over all
$\om'$ and the bound in Eq. \eq{zine} will lead to a divergence at small
$\om'$. To avoid this divergence, we need to investigate the bound in Eq.
\eq{inex} more carefully. Denote by $G(\om')$ the integral in Eq. \eq{inex},
i.e.
\bean
G(\om')=\int_0^{\infty}
\frac{e^{-x}e^{-\frac{2 M^2\om'(\omo-\ep)}{x}}}{|\ep \tilde u(x)|^k} dx.
\label{gom} 
\eean
First, rewrite $|\tilde u(x)|^2$ in Eq.\eq{abu} as
\bean
|\tilde u(x)|^2=\left(F_n(\om')+ 4 M\ln x\right)^2+\left(2M\pi+\frac{2
M^2\om'}{x} \right)^2,   \label{rabu}
\eean
where 
\bean
F_n(\om')\equiv 2 n\pi-4 M\ln\om'.
\eean
 In the following discussion, we choose $\Omega\ll 1$ and consider the
frequency range $\om'\leq\Omega$. For $n\gg 1$, 
$F_n(\om')$ is a large positive quantity. Substitute Eq. \eq{rabu}
into Eq. \eq{gom}
\bean
G(\om')=\int_0^{\infty}
\frac{e^{-x}e^{-\frac{2 M^2\om'(\omo-\ep)}{x}}}{\ep^k
\left[\left(F_n(\om')+ 4 M\ln x\right)^2+\left(2M\pi+\frac{2 
M^2\om'}{x} \right)^2   \right]^{k/2}} dx.
\label{gomn} 
\eean
I will evaluate the bound for  this integral in three domains of
$x$. Choose $a$ such that $a\ll1$ and $a n\gg 1$.  \\ \\ 
{\bf (1)}$D1=\{e^{-a F_n(\om')/4M}\leq x \leq e^{a
F_n(\om')/4M} \ 
and \ x \geq\frac{2 M^2\om'}{a F_n(\om')}\}$ \\
The integral in this
interval is approximately
\bean
G(\om', D1)=\int_{D1}
\frac{e^{-x}e^{-\frac{2 M^2\om'(\omo-\ep)}{x}}}{\ep^k
\left[F_n(\om') \right]^{k}} dx.
\eean
Since $D1$ is a subset of $(0,\infty)$ and the integrand is always
positive, we have
\bean
G(\om', D1)&\leq& \frac{1}{\left[\ep F_n(\om') \right]^{k}  } \int_{0}^{\infty}
e^{-x}e^{-\frac{2 M^2\om'(\omo-\ep)}{x}} dx \nonumber \\
&=&  \frac{ 2 \sqrt 2}{\left[\ep F_n(\om') \right]^{k}  }  
\sqrt{M^2(\omo-\ep)\om'} K_1\left( 2 \sqrt 2
\sqrt{M^2(\omo-\ep)\om'} \right) \label{gbk}
\eean
where $K_1$ is a modified Bessel function. For $z\rightarrow
0$, we have $K_1(z)\rightarrow \frac{1}{z}$ \cite{bes}. Therefore, for small
$\om'$, \eq{gbk} takes the form
\bean
G(\om', D1)\leq \frac{1}{\left[\ep F_n(\om') \right]^{k}  }. \label{gd1} 
\eean
{\bf (2)} $  D2=\{x\leq \frac{2M^2\om'}{a F_n(\om')} \}$  \\
The integral in this range is
\bean
G(\om',D2)=\int_0^{\frac{2 M^2 \om'}{a F_n(\om')}}
\frac{e^{-x}e^{-\frac{2 M^2\om'(\omo-\ep)}{x}}}{|\ep \tilde u(x)|^k} dx.
\label{go2} 
\eean
Note that  $a F_n(\om')\leq \frac{2 M^2\om'}{x}$ gives
\bean
e^{-\frac{2 M^2\om'(\omo-\ep)}{x}}\leq e^{-a F_n(\om') (\omo-\ep)}.
\label{eoi} 
\eean
Replacing $e^{-x}$ by $1$ and using Eqs. \eq{geqn} and \eq{eoi} to replace
the corresponding terms in the integrand of Eq. \eq{go2}, we obtain the
bound for $G(\om',D2)$,
\bean
G(\om',D2)\leq \frac{1}{(\ep 2 n\pi)^k} e^{-a
F_n(\om')(\omo-\ep)}\frac{2  M^2 \om'}{a F_n(\om')}. \label{gotb}
\eean
{\bf (3)} $D3\equiv\{x\in (0,\infty)|x\notin D1 \cup D2\} $ \\
Obviously, $D3$ is a subset of all positive $x$ satisfying $a
F_n(\om')\leq |4M \ln x| $. Then, it follows from Eq. \eq{gom} that 
\bean
G(\om', D3)\leq \int_0^{e^{-\frac{a F_n(\om')}{4M}}}
\frac{e^{-x}e^{-\frac{2 M^2\om'(\omo-\ep)}{x}}}{|\ep \tilde u(x)|^k} dx
+\int_{e^{\frac{a F_n(\om')}{4M}}}^{\infty}
\frac{e^{-x}e^{-\frac{2 M^2\om'(\omo-\ep)}{x}}}{|\ep \tilde u(x)|^k} dx.
\eean
Replace $|\tilde u(x)|$ by $2 n\pi$ and  $ e^{-x} e^{-\frac{2
M^2\om'(\omo-\ep)}{x}}$ by $1$  in the first integral. Replace $|\tilde
u(x)|$ by $2 n\pi$ and  $ e^{-\frac{2
M^2\om'(\omo-\ep)}{x}}$ by $1$  in the second integral. Then we obtain
the bound for $G(\om', D3)$ 
\bean
G(\om', D3)\leq \frac{1}{(\ep 2n\pi)^k}e^{-\frac{a F_n(\om')}{4M}}+\frac{1}{(\ep
2n\pi)^k} e^{- e^{\frac{a F_n(\om')}{4M}} }. \label{gtb}
\eean
Combining \eq{gd1}, \eq{gotb} and \eq{gtb}, we have the bound for
$G(\om')$ for $\om'\leq \Omega$, where $\Omega \ll 1$, 
\bean
G(\om')&\leq& \frac{1}{\left[\ep F_n(\om') \right]^{k}  } +\frac{1}{(\ep 2
n\pi)^k} e^{-a F_n(\om')(\omo-\ep)}\frac{2  M^2 \om'}{a
F_n(\om')} \nonumber \\
&+& \frac{1}{(\ep 2n\pi)^k}e^{-\frac{a F_n(\om')}{4M}}+\frac{1}{(\ep
2n\pi)^k} e^{- e^{\frac{a F_n(\om')}{4M}} }. \label{gfi}
\eean
Since $F_n(\om')$ can be arbitrarily large, we only need to keep the
first term on the right-hand-side of Eq. \eq{gfi}. This fact reveals that
the integration in Eq. \eq{gomn} is approximated by taking away the
$x$-dependent 
terms in the denominator of the integrand.  Thus, the bound for
$|\hat z(\om')|$ in \eq{inex} at small $\om'$ is 
\bean
|\hat z(\om'<\Omega)|\leq\frac{\sqrt{\ep}C_k  e^{-2M\pi
(\omo-\ep)}}{\om'\sqrt{\beta\omo+\gamma\ep}} 
\frac{1}{\left[\ep F_n(\om') \right]^{k}  } .
\label{smb} 
\eean
For $\om' \geq \Omega$, we simply use the bound \eq{zine}. Then
\bean
&&|\hat z(\om'>\Omega)| \nonumber \\
&\leq &\frac{\sqrt{\ep}C_k  e^{-2M\pi
(\omo-\ep)}}{\om'\sqrt{\beta\omo+\gamma\ep}} 
\frac{ 2 \sqrt 2 }{\ep^k (2n\pi)^k} \sqrt{M^2(\omo-\ep)\om'} K_1( 2 \sqrt 2
\sqrt{M^2(\omo-\ep)\om'} )
 \label{zilo}
\eean

Now we are ready to compute the particle creation rates at late
times. It follows from Eqs. \eq{smb} and \eq{zilo} that Eq. \eq{npc} is
bounded by   
\bean
N_{n\ep}(\omo)&\leq& |t(\omo)|^2\frac{\ep
C_k^2 e^{-4M\pi \omo} }{(\beta\omo+\gamma\ep)\ep^{2k}}\left[\int_0^{\Omega} 
\frac{1}{\om'} \frac{1}{(2 n\pi-4
M\ln\om')^{2k}}d\om' \right. \nonumber  \\ 
&+& \left.  \frac{8}{n^{2k}}\int_{\Omega}^\infty
\frac{M^2(\omo-\ep)}{\om'} K_1^2(2\sqrt 2 \sqrt{M^2 (\omo-\ep)\om'})d\om'
\right]. \label{nne}
\eean
Evaluating the first integral gives \\
\bea
\frac{4 M}{(2 k-1)}\frac{1}{(2 n\pi -4 M\ln \Omega)^{2k-1}}\approx
\frac{4 M}{(2 k-1)}\frac{1}{(2 n\pi)^{2k-1}} 
\eea

The second integral in Eq. \eq{nne} is convergent since $K_1(z)\sim
\sqrt{\frac{\pi}{2 z}}e^{-z}$ for large $|z|$ \cite{bes}. Since $n$
represents time, we conclude 
that the particle creation rate for any mode decays with time faster
than any power law. Furthermore, by summing over $n$ from any positive
integer to infinity, the bound in Eq. \eq{nne} is still finite. This means
that, as mentioned in the introduction,  the accumulation of particles
after an infinite time is finite.
This conclusion is derived from a particular Kruskal
transformation \eq{gku}, which corresponds to a particular process of
collapse.  A different process of collapse will give rise to a
different Kruskal coordinate $U'$, which is a smooth function of $U$. We
conjecture that our conclusion that the particle creation rate decays
with time faster than any power law is independent of the choice of the
Kruskal extension, i.e., independent of the details of collapse. As 
evidence in this conjecture, one can check, following similar steps,
that the smooth extension \eq{nkru} also gives the same result.  

\section{Stress energy tensor}
In two-dimensional Minkowski spacetime, the renormalized energy flux  in
a spacetime with a moving mirror boundary in the ``in'' vacuum state is
\cite{dfp}  
\bean
< T_{uu}>_q =\frac{1}{4\pi}\left[\frac{1}{4}\left(\frac{p''}{p'}\right)^2-
\frac{1}{6}\frac{p'''}{p'}\right], \label{stp}
\eean 
where $p=v=p(u)$ is the trajectory of the mirror. According to the
discussion in the last section, $p$ is exactly a Kruskal extension $U$
which describes a particular collapse.  The corresponding trajectory of
Kruskal extension \eq{gku} is thereby
\bean
u=-4 M \ln(-p)-\frac{2M^2}{p}. \label{mtr}
\eean
Straightforward calculation yields 
\bean
<T_{uu}>=\frac{p^3(p-2M)}{48\pi M^2(M-2p)^4},  
\eean
which approaches 
\bean
<T_{uu}>\sim\frac{1}{u^3} \label{tlu}
\eean
for large $u$. 
Therefore,  $<T_{uu}>$ decays as $1/u^3$.

The quantity  $<T_{uu}>$ can be estimated from the particle flux by
adding up all frequency modes at certain time 
\bean
<T_{uu}>\sim\int_0^\infty N_{n\ep}(\omo)\omo d\omo. \label{tnr} 
\eean
This formula, as discussed in \cite{dfp}, is a naive energy-particle
relation. It is correct only when particles emitted in different modes
are not correlated. However, it is worth comparing this naive $<T_{uu}>$
with the one in Eq. \eq{tlu}. If the naive one is smaller, it may indicate
some serious problems in our calculation of $ N_{n\ep}(\omo)$.
From the estimation of $ N_{n\ep}(\omo)$ in the last section, Eq. \eq{tnr}
suggests that  the naive $<T_{uu}>$ also should decay to zero faster than any
inverse power of $u$. So the result \eq{tlu} may seem to be
inconsistent with the particle creation results.
However, in the last section, we treated $\omo$ and $\ep$ as fixed while
allowing $n$ to be arbitrarily large. But Eq. \eq{tnr} requires us  to sum over
all modes at a fixed time $n$. So we need to reevaluate the bound of
$N_{n\ep}$ for arbitrarily small $\omo$. In this subsection, we focus on
the 
(1+1)-dimensional RN black  hole formed by collapse because our purpose is
to check the consistency of \eq{tlu} which is computed in a
1+1-dimensional spacetime. The calculation will be parallel to our 
four dimensional case in the previous sections. One important difference is
that the transmission amplitude $t(\omo)$ has unit magnitude  due to the fact
that a (1+1)-spacetime is conformal to Minkowski spacetime and therefore the
outgoing wave packet will not be scattered  when it is propagated
backward in time. So the bound in Eq. 
\eq{nne} still holds for the (1+1)-dimensional case except
$|t(\omo)|=1$. However, for our present purposes,  the bound in Eq.
\eq{inex} becomes 
inappropriate since $|\ep\tilde u(x)|\gg 1$ is not always true for
arbitrarily small $\ep$. So we stick to Eq. \eq{oni} and follow  similar
arguments. Denote by $H(\om')$ the integral  in \eq{gom}, i.e., 
\bean
H(\om')=\int_0^   
{\infty}\frac{e^{-x}e^{-\frac{2 M^2\om'(\omo-\ep)}{x}}}{(1+|\ep \tilde
u(x)|)^k} dx. \label{ng}
\eean
Consequently, the corresponding changes in Eq. \eq{gfi} become
\bean
H(\om')&\leq& \frac{1}{\left[1+\ep F_n(\om') \right]^{k}  } +\frac{1}{(1+\ep 2
n\pi)^k} e^{-a F_n(\om')(\omo-\ep)}\frac{2  M^2 \om'}{a
F_n(\om')} \\ \nonumber
&+& \frac{1}{(1+\ep 2n\pi)^k}e^{-\frac{a F_n(\om')}{4M}}+\frac{1}{(1+\ep
2n\pi)^k} e^{- e^{\frac{a F_n(\om')}{4M}} }. \label{cgo}
\eean
The last two terms in the bound are still negligible due to the
exponentials. Unlike before, we are not certain whether the second term
is much smaller than the first term since $\omo-\ep$ can be arbitrarily
small now. So we 
keep both of the terms. In order to perform integrals easily, we replace the
exponential and $\om'$ in the numerator by $1$ in the second term. Thus, 
\bean
 H(\om')&\leq& \frac{1}{\left[1+\ep F_n(\om') \right]^{k}  } +\frac{1}{(1+\ep 2
n\pi)^k} \frac{2  M^2 }{a F_n(\om')}  \label{fgo}
\eean
Then \eq{oni} becomes
\bean
|\hat
z(\om'<\Omega)|&\leq&\frac{\sqrt{\ep}C_k  e^{-2M\pi
    (\omo-\ep)}}{\om'\sqrt{\beta\omo+\gamma\ep}}\left( 
\frac{1}{\left[1+\ep F_n(\om') \right]^{k}  } +\frac{1}{(1+\ep 2 
n\pi)^k} \frac{2  M^2 }{a F_n(\om')}\right). \label{zio}
\eean
The modification for $\hat z(\om'>\Omega)$ in Eq. \eq{zilo} is
\bean
&&|\hat z(\om'>\Omega)| \nonumber \\
&\leq &\frac{\sqrt{\ep}C_k  e^{-2M\pi
    (\omo-\ep)}}{\om'\sqrt{\beta\omo+\gamma\ep}} 
\frac{ 2 \sqrt 2 }{(1+\ep 2 n\pi)^k} \sqrt{M^2(\omo-\ep)\om'} K_1\left(
2 \sqrt 2 
\sqrt{M^2(\omo-\ep)\om'} \right) \label{lob}
\eean

Thus, the bound on the particle number from \eq{npc} is
\bean
&&N_{n\ep}(\omo) \nonumber \\
&\leq& \frac{2\ep C_k^2  e^{-4M\pi (\omo-\ep)}}{\beta \omo+\gamma \ep}
\int_{0}^{\Omega} 
\frac{1}{\om'} \left(
\frac{1}{\left[1+\ep F_n(\om') \right]^{k}  }\right)^2
+\frac{1}{\om'}\left(\frac{1}{(1+\ep 2 n\pi)^k} \frac{2  M^2 }{a
F_n(\om')}\right)^2 d\om' \nonumber \\
&+&\frac{\ep C_k^2  e^{-4M\pi
(\omo-\ep)}}{\beta\omo+\gamma\ep}\frac{8}{(1+\ep 2 
n\pi)^{2k}}\int_{\Omega}^{\infty} \frac{M^2
(\omo-\ep)\om'}{\om'}K_1^2\left(2\sqrt 2  
\sqrt{M^2 (\omo-\ep)\om'}\right) d\om',  \nonumber \\
&& \label{nnp}
\eean 
where we have used the inequality $(A+B)^2\leq 2(A^2+B^2)$. The
integration over $(0,\Omega)$ gives
\bean
\frac{2\ep C_k^2  e^{-4M\pi (\omo-\ep)}}{\beta \omo+\gamma \ep}
\left(\frac{1}{(2 k-1)4 M\ep}\frac{1}{(1+2n\pi\ep)^{2k-1}}+\frac{M^3}{2
n\pi 
a^2(1+2n\pi\ep)^{2k}}\right), \label{bzd}
\eean
where we have neglected the $\Omega$-dependent term since it is small compared
with $n$. 
To estimate  the integration over $(\Omega,\infty)$, we first change
the integration variable from $\om'$ to $x=(\omo-\ep)\om'$. Thus the integral
becomes $M^2\int_{\Omega( \omo-\ep)}^\infty K_1^2
\left(2\sqrt{2}\sqrt{M^2 x}\right) 
d x$. We then split the integral into two terms as 
\bean
M^2\int_{\Omega( \omo-\ep)}^b K_1^2 \left(2\sqrt{2}\sqrt{M^2 x}\right)
d x +M^2\int_{b}^\infty K_1^2 \left(2\sqrt{2}\sqrt{M^2 x}\right)
d x,
\eean
where $b>\Omega (\omo-\ep)$ and $bM^2\ll 1$. Thus, we can use the
approximate form of $K_1(z)\sim 1/z $ again for the first term. The
second term is just a 
constant and negligible compared to the first term when $\omo$ is taken
small enough. Therefore, the integral is approximately
$\frac{-M^2}{2\sqrt{2}} \ln [\Omega(\omo-\ep)]$. Together with a
coefficient, the second integral in Eq. \eq{nnp} yields
\bean
\frac{\ep C_k^2  e^{-2M\pi \omo}}{\beta\omo+\gamma\ep}\frac{8}{(1+\ep 2
n\pi)^{2k}}\frac{-M^2}{2\sqrt{2}} \ln [\Omega(\omo-\ep)].  \label{bip}
\eean 
As discussed above, the main contribution to $<T_{uu}>$ comes from the
low-frequency 
integration  $\int_{0}^\delta N(\omo)\omo d\omo$, where $\delta$ is a
small constant. Since it is assumed $\ep<\omo$, we can no longer treat
$\ep$ as constant when performing the integral. For convenience, we
choose $\ep$ to be a small fraction of $\omo$. It follows from Eqs. \eq{bzd}
and \eq{bip} that 
\bean
\int_{0}^\delta N_{n\ep}(\omo)\omo d\omo \leq \frac{K}{n}, \label{fb}
\eean
where $K$ is a constant. Thus, although the particle creation rate in
each individual mode goes to zero faster than any inverse power of time,
the energy flux estimated from the particle creation rate decays only as
$1/u$ on account of contributions from the very low frequency modes. 
The bound in Eq. \eq{fb} decays more slowly with time than Eq. \eq{tlu}. This
difference may be due to 
the fact that the bound \eq{fb} is not sharp enough. An alternative
possible explanation is that, as pointed out by Davies and
Fulling \cite{dfp}, 
the relationship between particle fluxes and energy fluxes can be more
subtle than \eq{tnr} as a result of destructive interference between
different modes. At this stage, we do not know if there exists
interference since we only have the upper bound on the number of particles. 

One can also ask if the $1/u^3$ decay rate of energy flux is
independent of the details of the collapse. Since each collapse
corresponds to a smooth extension, let us consider another smooth extension $q$
and the function $q=q(u)$, which gives the related energy flux, 
\bean
< T_{uu}>_q =\frac{1}{4\pi}\left[\frac{1}{4}\left(\frac{q''}{q'}\right)^2-
\frac{1}{6}\frac{q'''}{q'}\right]. \label{stq}
\eean 
Let $q=g(p)$. From $q(u)=g(p(u))$, we derive the relationship between
$ < T_{uu}>_q$ and  $ < T_{uu}>_p$ (the energy flux associated with $p$),
\bean
< T_{uu}>_q =<
T_{uu}>_p+[p'(u)]^2\frac{3g''(p)^2-2g'(p)g'''(p)}{48\pi g'(p)^2}.
\label{tpq}
\eean  
Since $p$ and $q$ are two smooth coordinates, the derivatives of
$g(p)$ must be finite and $g'(p)\neq 0$. From Eq. \eq{mtr},
$[p'(u)]^2$ goes as $1/u^4$ at late 
times. Therefore, the second term on the right-hand side of \eq{tpq}
is negligible compared to the first term which goes as
$1/u^3$. Therefore, we showed that the $1/u^3$ decay rate is invariant
under a smooth change of mirror trajectory. This fact agrees with the
fact that the
late-time radiation is  independent of the
details of collapse. If one applies the extension \eq{sgku} to compute
the energy flux, as done in \cite{three}, the energy flux would be
identical zero at late times. This is because the extension \eq{sgku}
fails to be a smooth one (see the comment at the end of section 2.2)
at late times (around $U=0$).  

\section{Conclusions}
We have calculated the number of particles  created from a massless scalar
field on an extremal RN black hole spacetime formed by collapse. We
found that for 
each mode associated with a wave packet, the rate of particle creation
drops off to zero 
faster than any inverse power of time at late times. Consequently, even
after an infinite time, the number of particles detected by a detector
sensitive to  a certain frequency is finite. This result
confirms that extremal black 
holes do not create particles. In the (1+1)-dimensional case, the
stress-energy flux falls off as $1/u^3$. This result is not in
contradiction with the much more rapid decay of each mode, since the
very low frequency modes do not achieve their  asymptotic decay rate
until very late times.
\\ \\
\noindent
\begin{center}
{\bf  \large Acknowledgments} 
\end{center}

It is my pleasure to express my thanks to professor Robert M. Wald for
many helpful discussions on this project. 

This research was supported  by NSF grants/PHY00-90138 to the
University of Chicago. This work was also supported in part by  a
grant, through CENTRA, from FCT (Portugal).

\appendix
\section*{Appendix: Proof of Lemma 1}
Let $R$ denote the radius of $C_2$ and
$I_R\equiv\int_{C_2}e^{icz}F(z)$. Substitute $z$ by $R e^{i\theta}$. Then
\bean
I_R=\int_{\frac{\pi}{2}}^{\frac{3\pi}{2}}e^{icR\cos\theta-cR\sin\theta}
F(Re^{i\theta})iRe^{i \theta}d\theta. 
\eean
Since $\lim_{|z|\rightarrow\infty}|F(z)|\rightarrow 0$, for any $\ep'>0$,
we can find $R$ 
such that
\bean
|I_R|&\leq& \ep' R\int_{\frac{\pi}{2}}^{\frac{3\pi}{2}}
e^{-cR\sin\theta}d\theta \nonumber \\
&\leq& \ep' R\int_{0}^{\frac{\pi}{2}}
e^{-cR\cos\theta}d\theta. 
\eean
In the range $[0,\frac{\pi}{2}]$,
\bea
\cos\theta\geq 1-\frac{2}{\pi}\theta
\eea
Therefore,
\bea
|I_R|&\leq& \ep' R\int_0^{\frac{\pi}{2}} e^{-cR(1-\frac{2}{\pi}
\theta)}d\theta \\ 
&\leq& \frac{\pi}{2c}\ep'(1-e^{-cR}).
\eea
Therefore,
\bean
\lim_{R\rightarrow\infty}|I_R|=0.
\eean

\end{document}